\documentclass[%
 reprint,
 superscriptaddress,
 amsmath,amssymb,
 aps,
]{revtex4-2}
\usepackage{graphicx}% Include figure files
\usepackage{dcolumn}% Align table columns on decimal point
\usepackage{bm}% bold math
\usepackage{subfig}
\usepackage{caption}
\usepackage{array}
\usepackage{booktabs}
\usepackage{multirow}
\usepackage{siunitx}
\usepackage[utf8]{inputenc}
\usepackage[T1]{fontenc}
\usepackage{mathptmx}
\usepackage{etoolbox}
\usepackage{hyperref}
\begin{document}

\preprint{APS/123-QED}

\title{Multi-beam-energy control unit based on triple bend achromats}% Force line breaks with \\

\author{Liuyang Wu}
\affiliation{Shanghai Advanced Research Institute, Chinese Academy of Sciences, Shanghai, China}
\affiliation{ShanghaiTech University, Shanghai, China}

\author{Zihan Zhu}
\altaffiliation{Present address: SLAC National Accelerator Laboratory, Menlo Park, CA 94025, USA}
\affiliation{Shanghai Institute of Applied Physics, Chinese Academy of Sciences, Shanghai, China}
\affiliation{University of Chinese Academy of Sciences, Beijing, China}

\author{Bingyang Yan}
\affiliation{Shanghai Institute of Applied Physics, Chinese Academy of Sciences, Shanghai, China}
\affiliation{University of Chinese Academy of Sciences, Beijing, China}
	
\author{Jiawei Yan}
 \email{jiawei.yan@xfel.eu}
\affiliation{European XFEL, Schenefeld, Germany}
	
\author{Haixiao Deng}
 \email{denghx@sari.ac.cn}
\affiliation{Shanghai Advanced Research Institute, Chinese Academy of Sciences, Shanghai, China}

\date{\today}% It is always \today, today,

\begin{abstract}
X-ray free electron lasers (XFELs) are the new generation of particle accelerator-based light sources, capable of producing tunable, high-power X-ray pulses that are increasingly vital across various scientific disciplines. Recently, continuous-wave (CW) XFELs driven by superconducting linear accelerators have garnered significant attention due to their ability to enhance availability by supporting multiple undulator lines simultaneously. However, different undulator lines typically require distinct electron beam qualities, particularly varying electron beam energy to achieve a wide range of photon energy tunability. Consequently, precise bunch-to-bunch control of electron beam energy is essential. A double-bend achromat based electron beam delay system has been proposed to enable multi-beam energy operations in CW-XFELs. In this paper, we introduce a novel delay system comprising four triple-bend achromats (TBAs). Based on parameters of the Shanghai High-Repetition-Rate XFEL and Extreme Light Facility, start-to-end simulations demonstrate that the TBA-based delay system achieves better electron beam qualities while providing a wide beam energy tuning range.
\end{abstract}

\maketitle

X-ray free electron lasers (XFELs) are capable of generating ultra-short, high-brightness X-ray pulses, which offer new opportunities for cutting-edge research in biomedicine, chemistry, physics, and materials science \cite{bostedt2016linac,Huang2021,kang2015crystal}. Different scientific experiments often have varying requirements for XFEL pulse characteristics, such as the photon energy and pulse length. To meet as many of these requirements as possible, it is common to employ multiple undulator lines at the end of the linear accelerator simultaneously \cite{Faatz_2016,PhysRevAccelBeams.19.020703}. 

The FEL resonance condition \cite{BONIFACIO1984373} follows
\begin{equation}
 \lambda=\frac{\lambda_{u}}{2\gamma^2}(1+\frac{K^2}{2})    
\end{equation}     
\noindent where $\lambda$ is the radiation wavelength, $\lambda_{u}$ is the length of the undulator period, $\gamma$ is the Lorentz factor of the energy of the electron beam, and K is the undulator strength parameter. The tuning range of the undulator strength typically limits the tunability of the photon energy range for a certain electron beam energy. As a result, to further increase the photon energy range of radiation light, it is critical to control the electron beam energy for each undulator line independently. There are two main approaches to realizing the multi-beam energy operation of an XFEL facility. One approach \cite{milne2017swissfel} is to utilize a switchyard line located in the linac to transport the relatively low-energy electron beam to a specific undulator line to generate X-ray pulses with lower photon energy while the other undulator lines cover higher photon energy. However, the branch line scheme still has a restricted photon energy tuning range for each undulator line. Moreover, it requires an additional beam transport line and more accelerator tunnel space. In addition, by adjusting the trigger frequency of specific RF units in the linac \cite{hara2013time} , different electron beams are either accelerated or not accelerated as they pass through these units, resulting in electron beams of varying energy at the end of the linac. This operation mode is highly flexible and does not degrade the electron beam quality but it cannot be applied to linacs operating in continuous-wave (CW) mode. 

\begin{figure*}[!htb]
   \centering
   \includegraphics*[width=\textwidth]{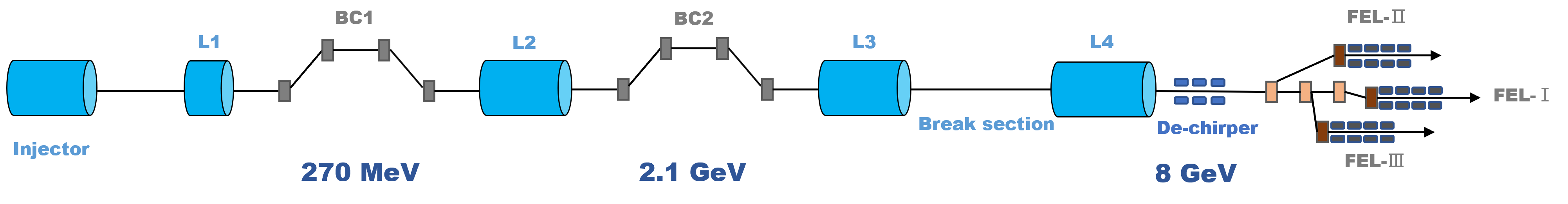}
   \caption{\label{fig:fig1}~Layout of SHINE.}  
\end{figure*}

The construction of high-repetition-rate XFEL worldwide facilities based on superconducting (SC) linac has been booming in recent years. The European XFEL \cite{decking2020mhz} has achieved a repetition rate of up to 4.5 MHz in burst mode since its first lasing in 2017. The LCLS-II \cite{bourzac2023world} achieved its first lasing as a CW XFEL facility capable of generating XFEL pulses at up to 1 MHz repetition. As the first hard X-ray FEL facility in China, Shanghai High-Repetition-Rate XFEL and Extreme Light Facility (SHINE) \cite{liu2023status} began construction in 2018 and is expected to achieve its first lasing in 2025. With a repetition rate of up to 1 MHz, SHINE employs multiple undulator lines to support the simultaneous operation of numerous experimental stations. The multi-beam-energy operation of a high-repetition-rate XFEL facility is critical to improve its usability. Recently, two methods have been proposed to realize multi-beam-energy control for SC linac. One method is based on the off-frequency detuning of the SC linac \cite{zhang2019multienergy, pfeiffer2022rf}, requiring mechanical perturbations of the cavity controlled by a frequency tuner. In addition, an electron bunch delay system \cite{PhysRevAccelBeams.22.090701} based on the double-bended achromats (DBAs) has been proposed. Such an energy control unit can achieve beam delay on a bunch-to-bunch basis. The system is placed in front of the last accelerator section of SHINE, changing the acceleration phase of the delayed electron beam in this section. As a result, electron beams with large energy differences can be obtained at the end of the linac based on the parameters of SHINE. Although the concept is promising, it is still in the preliminary stage since the electron beam qualities, especially the longitudinal phase space, could be affected by such a system.  In this study, we further extend the design of the multi-beam-energy operation. We propose an electron bunch delay system based on the triple-bend achromats (TBAs). Such a TBA-based electron beam delay system is designed to be isochronous and has good transverse dispersion control and emittance suppression. Numerical simulations show that this system can delay the electron bunch by more than 115 mm, half the RF period of the 1.3 GHz accelerated structure, without significantly degrading of the electron beam longitudinal phase space. In the following parts, the SHINE linac section, the design of the delay system based on TBAs, and the results of the start-to-end simulations will be presented. Finally, the conclusions will be summarized and possible future applications of the delay systems will be discussed.

SHINE employs a SC linac that generates high-brightness electron beams with a repetition rate of up to 1 MHz \cite{liu2023status} and an energy of up to 8 GeV. These electron beams are then separately transported into three undulator lines, enabling the output of XFEL pulses with photon energy covering the range from 0.4 to 25 keV. The layout of the SHINE facility is illustrated in Fig. \ref{fig:fig1}.

The SHINE injector is based on a VHF photocathode electron gun, which produces an electron beam with a repetition rate of 1 MHz, an energy of 100 MeV, and a charge of 100 pC. Following the injector is the main accelerator section, which consists of four accelerator sections and two magnetic bunch compressors. In the first accelerator section (L1), the electron beam is accelerated in the off-crest phase to form an energy chirp in the longitudinal phase space. After passing through the 3.9 GHz high-frequency cryomodules, the high-order energy chirp is compensated, which is used for the subsequent compression of the bunch. Then the electron beam with an energy of 270 MeV is compressed after passing through the first magnetic compressor. The second accelerator section (L2) is located between the first and second magnetic chicane, accelerating the electron beam to 2.1 GeV. After the second magnetic compression section (BC2), the third and fourth accelerator sections (L3 and L4) further accelerate the electron beam to 8 GeV and the peak current intensity reaches around 1500 A. A metal corrugated structure \cite{PhysRevLett.113.254802} located at the end of the linac section is designed to compensate for the energy chirp of the electron beam. Following the main acceleration section, the beam distribution system distributes the electron beam to three undulator lines (FEL-I, FEL-II, and FEL-III). FEL-I employs a permanent magnet planar undulator with a period of 26 mm, covering a photon energy range of 3-15 keV, and operates in spontaneous self-amplified emission (SASE) \cite{kondratenko1980generating,bonifacio1984instabilities} and self-seeding modes \cite{Geloni20092011,amann2012demonstration}. FEL-II covers a photon energy range of 0.4-3 keV, and has several operation modes including SASE, self-seeding, high-gain harmonic generation (HGHG) \cite{yu2000high,allaria2012highly}, and echo-enabled harmonic generation (EEHG) \cite{stupakov2009using,zhao2012first} modes, and the cascaded EEHG-HGHG \cite{YANG2022167065} mode. FEL-III employs an SC undulator with a period of 16 mm, covering a photon energy range of 10-25 keV, which also operates in SASE and self-seeding modes. 

In this study, we consider placing an electron bunch delay system in the break section between the L3 and L4 to delay the electron beam before the L4 acceleration section. This would result in a change to the acceleration phase experienced by the bunches within the L4 section, thereby producing electron bunches with a large energy difference. Accordingly, the delay distance provided by this system must be adaptively regulated between zero and one-half of the RF period to achieve the optimal electron beam energy control range. In the case of the 1.3 GHz RF acceleration section, the length of half of the RF period is 115.4 mm.

 \begin{figure}[!htb]
   \centering
   \captionsetup{justification=raggedright, singlelinecheck=false}
   \includegraphics[width=0.45\textwidth]{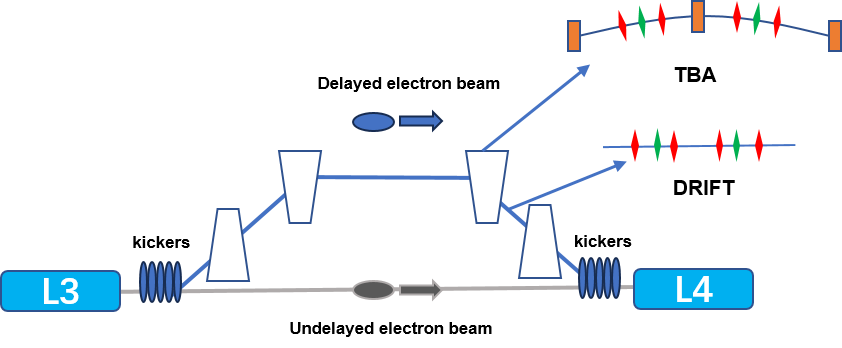}
   \caption{\label{fig:fig2}~Layout of the electron beam delay system based on TBAs. The orange rectangle represents a dipole magnet and the rhombus represents a quadrupole. The green and red rhombuses represent focusing and defocusing quadrupoles, respectively.}
\end{figure}

The proposed electron beam delay system is designed to be composed of four TBAs with horizontal bends, as shown in Fig. \ref{fig:fig2}. To minimize the impact on beam quality, the delay system should be achromatic and isochronous. To realize isochronism, we need to precisely optimize the lattice and control the $R_{56}$ value of the whole system to the micrometer level. At the entrance and exit of the delay system, two sets of vertical kickers are placed to separate the electron beams that need to be delayed from the normally accelerated electron beams. This allows for precise delay of certain electron bunches and enables bunch-to-bunch control. The design of these two sets of vertical kickers is similar to that of the DBA-based energy control unit \cite{PhysRevAccelBeams.22.090701}, with a total deflection angle of the septum magnets of 1 mrad and a drift space behind the septum magnet of 15 m, which separating the deflected and undeflected electron beams by 17.30 mm. The TBA-based delay system not only provides a large electron beam delay but also minimizes the degradation of the beam quality. 

The most critical issue of this TBA-based electron beam delay system is how to avoid the growth of the normalized emittance of the electron beam due to the coherent synchrotron radiation (CSR) effect. To minimize the effect of CSR, an optical balance method \cite{di2013cancellation} is used in the design of the lattice. To achieve this, each TBA contains six quadrupole magnets, and the drift section between two consecutive TBAs also contains six quadrupole magnets. The phase advance between two consecutive dipole magnets inside the TBA is controlled to be $\pi$. We use the non-dominated sorting genetic Algorithm II (NSGA-II) \cite{deb2000fast} to optimize the strengths of these quadrupole magnets magnets as well as the length of the drift sections between the magnets, ensuring that the system is both achromatic and isochronous.
The delay distance induced by one TBA can be estimated as

\begin{equation}
\Delta Z_{TBA}=\frac{L1}{2}(\frac{1}{cos|\theta|}-\frac{1}{cos|2\theta|}-2)  
\end{equation}

\noindent where L1 is the projection length of the TBA on the longitudinal axis, $\theta$ is the dipole angle of the TBA, The delay distance induced by an inclined drift space between two consecutive TBAs can be calculated as

\begin{equation}
\Delta Z_{Drift}=L2(\frac{1}{cos|3\theta|}-1)  
\end{equation}

\noindent where L2 is the projection length of the drift space on the longitudinal axis. The delay distance induced by the whole delay system is

\begin{equation}
\Delta Z=4\Delta Z_{TBA}+2\Delta Z_{Drift}   
\end{equation}

\begin{figure}[!htb]
\centering
\subfloat{
		\includegraphics[scale=0.35]{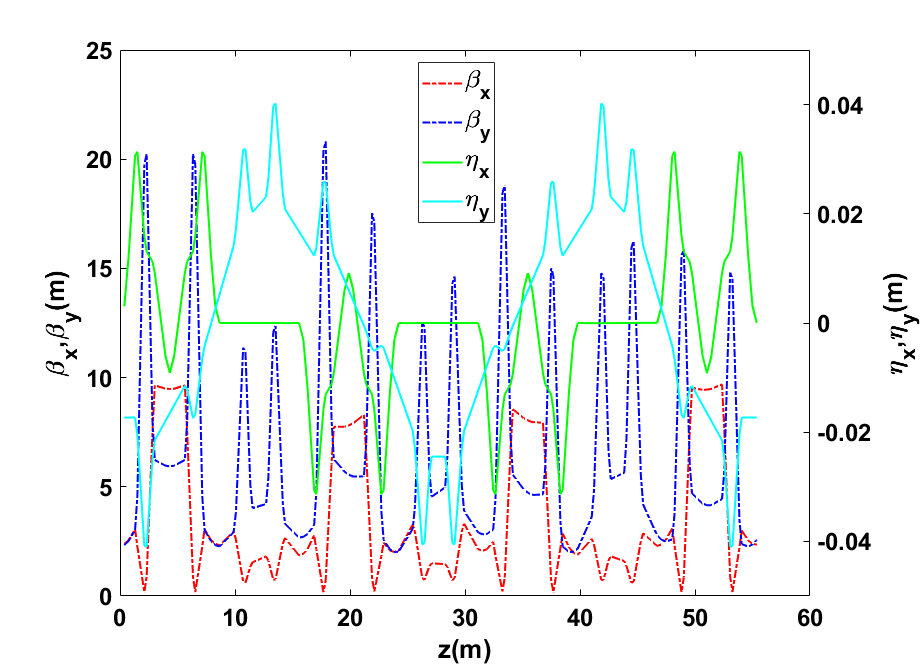}}\\
\subfloat{
		\includegraphics[scale=0.25]{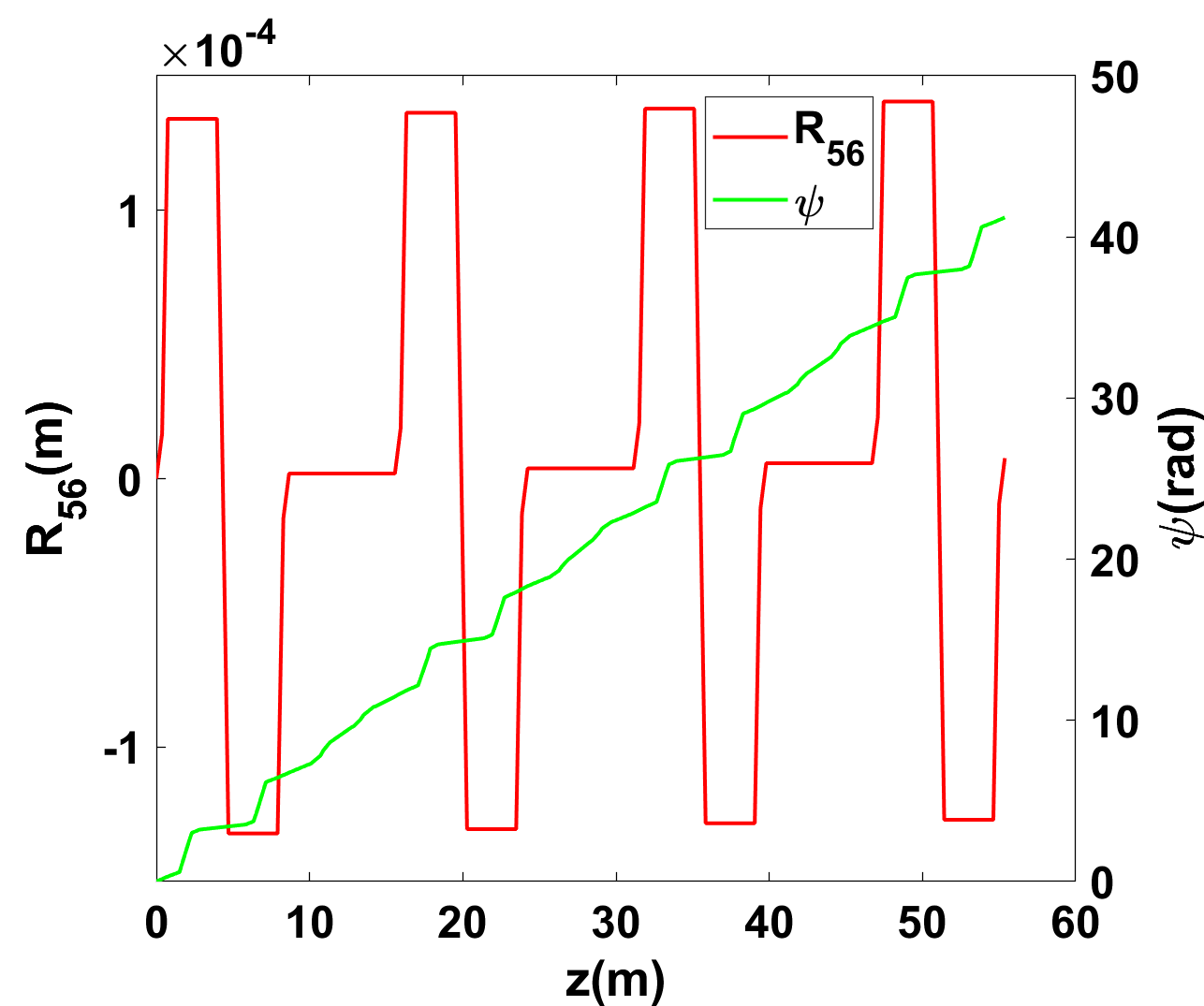}}
\captionsetup{justification=raggedright, singlelinecheck=false}
\caption{\label{fig:fig3}~Upper: Courant-Snyder parameters $\beta$ function, and $\eta$ function along the TBA-based electron bunch delay system. Bottom:$R_{56}$ and $\psi$ function along the TBA-based electron bunch delay system.}
\end{figure}

Simulations of beam transport using ELEGANT \cite{borland2000flexible} enable more accurate calculation of delay distances. To achieve a delay distance of up to 115.4 mm, L1, L2 and $\theta$ are set to 8.65 m, 6.57 m and 1.876$^\circ$, respectively. The delay distance can be effectively controlled by adjusting the angle of the dipole magnets. Figure. \ref{fig:fig3} illustrates the lattice design of the $\beta$ function, the dispersion function, $R_{56}$, and the horizontal phase advance $\psi$ along the delay system when the dipole angle is set to 1.876$^\circ$

\begin{figure}[h]
   \centering
   \includegraphics[width=0.45\textwidth]{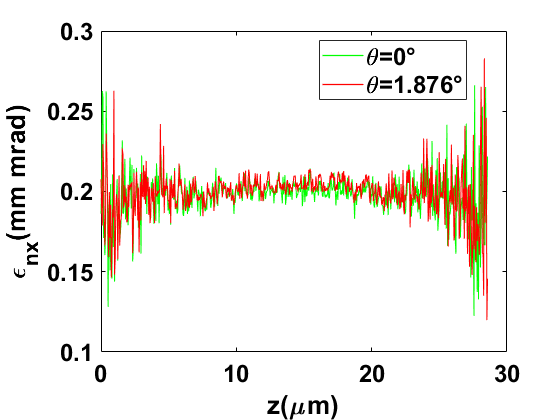}
   \captionsetup{justification=raggedright, singlelinecheck=false}
   \caption{\label{fig:fig4}~Slice normalized horizontal emittance of the Gaussian electron beam after passing through the delay system.}  
\end{figure}

To further validate the optimized lattice, a 100 pC Guassian distribution electron beam is fed into the delay system with the dipole magnets set at angles of 0$^\circ$ and 1.876$^\circ$, respectively. The normalized horizontal emittance of the electron beam at the entrance of the delay system is set to 0.2 mm$\cdot$mrad, the relative energy spread is 0.1\%, the bunch length is 10.8 $\mu$m (FWHM), and the peak current is 2.5 kA. CSR effect and the space-charge effect were considered when ELEGANT was used for tracking simulations with one million macroparticles. The normalized horizontal emittance at the end of the delay system is 0.2 and 0.21 mm$\cdot$mrad when the bend angle is set to 0$^\circ$ and 1.876$^\circ$, respectively. Figure. \ref{fig:fig4} illustrates the slice normalized horizontal emittance of this Gaussian electron beam after passing through the delay system. The simulation result demonstrates that the growth of the normalized horizontal emittance can be well controlled at the maximum adjustable angle.

\begin{table*}
    \caption{\label{tab:table1} Simulation results of electron beam of double-horn current profile in different delay system.}
    \begin{ruledtabular}
        \begin{tabular}{c|ccc}
            & Delay distance (mm) & Energy (GeV) & \(\epsilon_{nx}\) (mm$\cdot$mrad)\\
            \hline
            \multirow{4}{*}{DBA/TBA}
            & 0 & 8.74 & 0.22 / 0.22\\
            & 57.7 & 5.11 & 0.24 / 0.23\\
            & 76.9 & 3.30& 0.28 / 0.23 \\
            & 115.4 & 1.48 & 0.33 / 0.25 \\     
        \end{tabular}
    \end{ruledtabular}
\end{table*}

\begin{figure}[!htb]
\centering
\subfloat[\centering]{
    \includegraphics[scale=0.25]{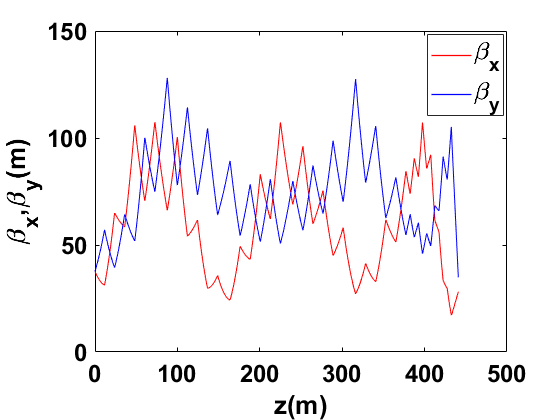}}
\subfloat[\centering]{
    \includegraphics[scale=0.25]{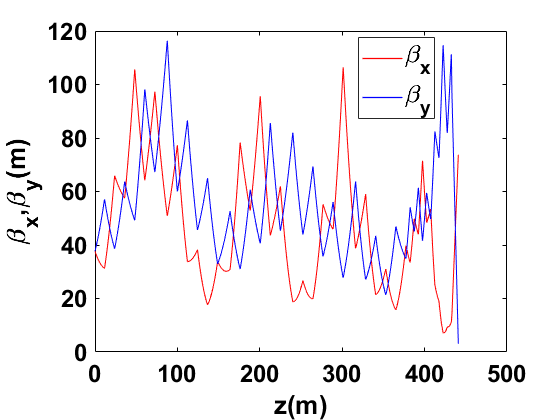}}
\\
\subfloat[\centering]{
    \includegraphics[scale=0.25]{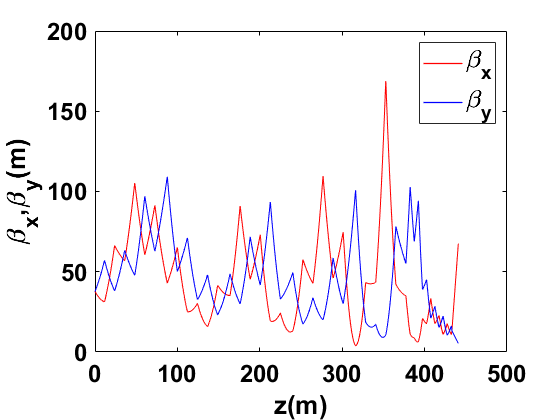}}
\subfloat[\centering]{
    \includegraphics[scale=0.25]{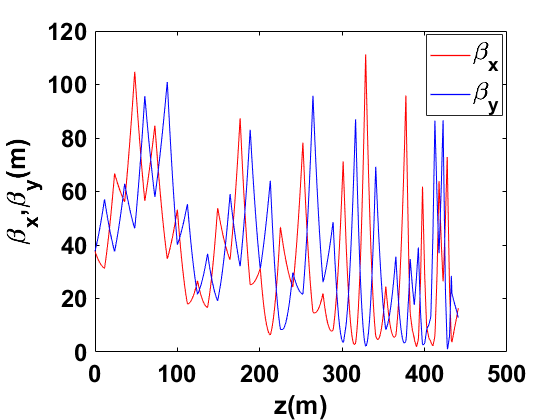}}
   \captionsetup{justification=raggedright, singlelinecheck=false}
   \caption{\label{fig:5}~$\beta$ functions along optimized L4 and dechirper section, figure a, b, c, and d correspond to different electron beam energy of 8.0, 4.68, 3.03, and 1.39 GeV at the end of the linear section, respectively.}  
\end{figure}

We further employ start-to-end simulations using SHINE parameters to analyze the delay system. We firstly tracked the main accelerator and electron bunch delay system sections using the electron bunch with a double-horn current profile, which is the baseline deign of SHINE in the previous work \cite{PhysRevAccelBeams.22.090701}. The energy of the electron beam at the exit of L3 section is 3.3 GeV, and the energy of the electron beam at the end of the linac can be controlled from 1.48 to 8.74 GeV with the utilization of the delay system. For this study, we adopted the lattice configurations of the L4 and dechirper sections, previously optimized for multi-beam-energy operation in \cite{PhysRevAccelBeams.22.090701}. The collective effects, such as the CSR effect, longitudinal space charge effect, and wakefield effect, are considered in the simulations. The dipole angle needs to increase accordingly and the delay distance increases, which leads to an increase in the growth of the emittance due to the CSR effect. Compared to the DBA-based electron bunch delay system, the normalized beam emittance growth in the TBA delay system is only 50\%, 16.7\% and 27.3\% of those in the DBA-based delay system, when the delay distance is 57.7 mm, 76.9 mm and 115.4 mm respectively. The specific results are shown in Table \ref{tab:table1}. With the values of $\beta_{x}$ are of the same order of magnitude, this TBA-based electron bunch delay system has one-tenth the maximum value of $\beta_{y}$ of the dipoles in DBA-based electron bunch delay system. In addition, at the same delay distance, the deflection angle of one dipole magnet in the TBA is smaller, which also benefits the suppression of normalized horizontal emittance growth. 

Subsequently, we used another working point \cite{zhu2022inhibition}, the electron bunch of which is optimized to a flat-top current profile to enhance the lasing performance of the cascaded EEHG-HGHG scheme. The energy of the electron beam at the exit of L3 section is 3.03 GeV, and the energy of the electron beam at the end of the linac can be controlled from 1.39 to 8.0 GeV with the utilization of the delay system. In the L4 and dechirper sections, the quadrupole magnets were originally designed to accommodate a maximum acceleration energy of 8.0 GeV. Similar to the previous working point, the inclusion of the delay system requires the lattice design of the L4 and dechirper sections to withstand a wide energy range, extending down to 1.39 GeV. Without adjusting the lattice of L4 and dechirper sections, the value $\beta_{x}$ at the end of the linac at the lowest beam energy can exceed 10 million, which is unacceptable. Therefore, we also optimized the strength of the 41 quadrupole magnets in the L4 and dechirper sections for this working point using the NSGA-II algorithm. The $\beta$ function for different electron beam energy with the optimized lattice is shown in Fig. \ref{fig:5}. The results demonstrate that the optimized lattice can effectively control the envelope of electron bunches with different beam energy. 

\begin{table*}
    \caption{\label{tab:table2} Simulation results of electron beam of flat-top current profile in different delay system.}
    \begin{ruledtabular}
        \begin{tabular}{c|cccc}
            & Delay distance (mm) & Energy (GeV) & Peak current (kA) & $T_{566}$ (mm)\\
            \hline
            \multirow{4}{*}{DBA/TBA}
            & 0 & 8.0 & 1.68 / 1.68 & 0 / 0\\
            & 57.7 & 4.68 & 2.21 / 1.73& 59.80 / 10.13 \\
            & 76.9 & 3.03 & 2.86 / 1.76& 79.75 / 13.49\\
            & 115.4 & 1.39 & 5.25 / 1.83& 119.76 / 20.24\\
        \end{tabular}
    \end{ruledtabular}
\end{table*}

\begin{figure}[!htb]
\subfloat[\centering]{
    \includegraphics[scale=0.25]{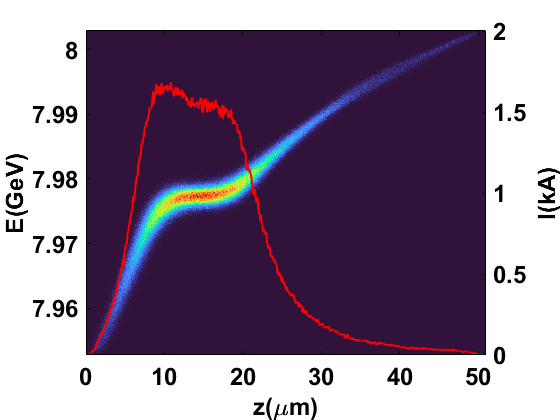} 
    \label{fig:fig6_a}}  
\subfloat[\centering]{
    \includegraphics[scale=0.25]{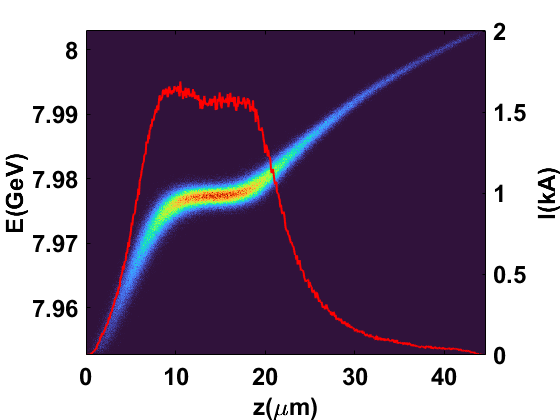}
    \label{fig:fig6_b}}  
\\
\subfloat[\centering]{
    \includegraphics[scale=0.25]{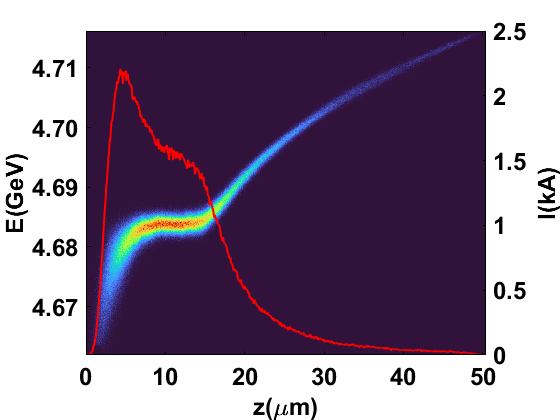}
    \label{fig:fig6_c}}
\subfloat[\centering]{
    \includegraphics[scale=0.25]{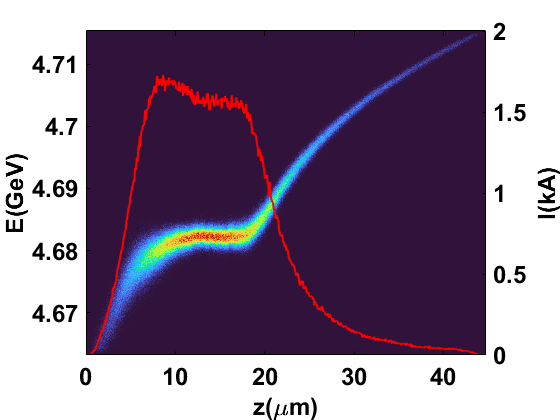}
    \label{fig:fig6_d}}
\\
\subfloat[\centering]{
    \includegraphics[scale=0.25]{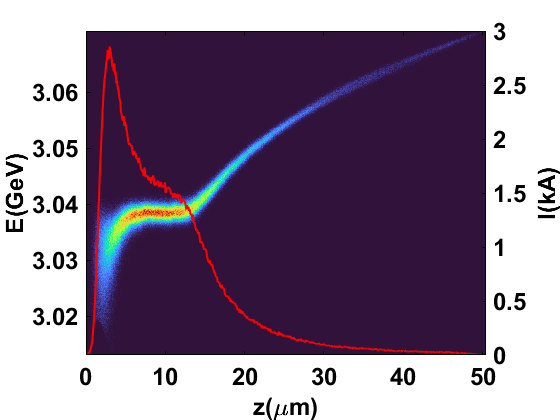}
    \label{fig:fig6_e}}
\subfloat[\centering]{
    \includegraphics[scale=0.25]{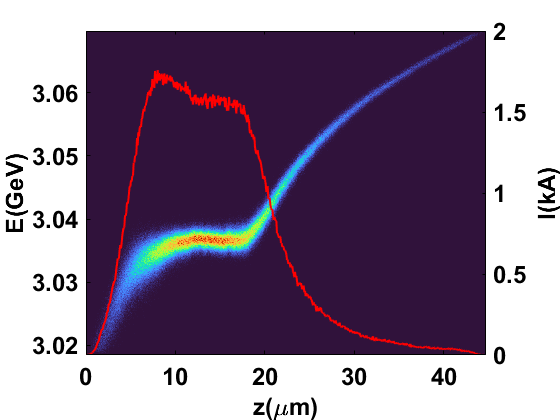}
    \label{fig:fig6_f}}
\\
\subfloat[\centering]{
    \includegraphics[scale=0.25]{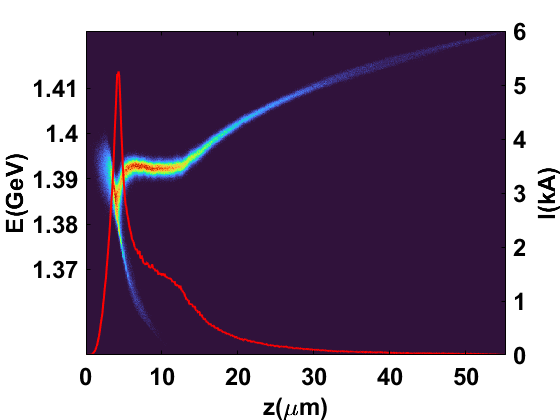}
    \label{fig:fig6_g}}  
\subfloat[\centering]{
    \includegraphics[scale=0.25]{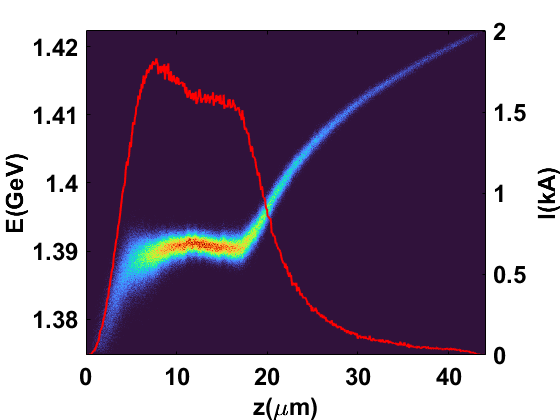}
    \label{fig:fig6_h}}
\captionsetup{justification=raggedright, singlelinecheck=false}
\caption{\label{fig:6}~Longitudinal phase spaces and the corresponding current distributions of the electron bunch 
        at the end of the linac when the DBA-based (left) and TBA-based delay system (right) is employed, respectively. The delay distance from top to bottom is 0, 57.7, 76.9 and 115.4 mm, respectively.}
\end{figure}

The simulations show that the TBA-based electron bunch delay system holds superior performance in maintaining the phase space while suppressing the normalized emittance growth when the working point with a flat-top current profile is employed. At delay distances of 0, 57.7, 76.9, and 115.4 mm, the normalized horizontal emittance at the exit of the delay system is 0.22, 0.26, 0.27, and 0.29 mm$\cdot$mrad for the TBA-based delay systems, compared to 0.22, 0.23, 0.24, and 0.35 mm$\cdot$mrad for the DBA case. Simulation results of four typical electron beams through the DBA-based and TBA-based delay systems are summarized in Table \ref{tab:table2}. The increase in the normalized horizontal emittance at the maximum delay distance of the DBA-based delay system is around 20.7\% larger than the TBA-based delay system. More importantly, the TBA system minimizes longitudinal phase space distortion, even at large delays. As shown in Fig. \ref{fig:6}, the longitudinal phase space and current profiles of electron beams passing through the TBA system remain largely undistorted across all delay distances. In contrast, the DBA system introduces significant distortion in both phase space and current distribution as the delay increases, with the peak current exceeding 5000 A at 115.4 mm delay (Fig. \ref{fig:fig6_g}). This distortion in the DBA case arises primarily from the large energy chirp of the electron beam, which is threefold higher than that of the double-horn current profile working point. For a beam with small transverse emittance, the transformation of its distribution in longitudinal bunch coordinate $z$ is dominated by the first- and second-order transport matrix elements $R_{56}$ and $T_{566}$ \cite{PhysRevAccelBeams.24.060701,england2005sextupole}:
\begin{equation}
 z =  z_{i} + R_{56}\delta_{i}+ T_{566}\delta_{i}^2 +O(\delta_{i}^3),
\end{equation}
where $\delta_{i}$ is the momentum deviation, $z_{i}$ is the longitudinal bunch coordinate in the beamline entrance.
While both delay systems are isochronous ($R_{56}$ controlled), the TBA system achieves superior $T_{566}$  management. This capability is critical for mitigating nonlinear phase space distortions, especially under the amplified influence of $T_{566}$ caused by the larger energy chirp in the flat-top profile case.

In this study, we proposed a TBA-based beam delay system to largely improve the photon energy tunability of each undulator line for a CW-XFEL. This delay system is capable of changing the accelerating phase of the electron beam bunch in the final acceleration section without compromising beam quality. When the electron beam is controlled in the range of 1.39-8.0 GeV, the transverse emittance growth is suppressed, and the longitudinal phase space of the electron beam is well maintained. Combined with the faster kickers and septum, we can realize the bunch-to-bunch energy control in a CW-XFEL. Such a large range of electron beam energy control will allow each undulator line to cover both soft and hard X-rays. Compared to the previously proposed DBA-based electron bunch delay system, the TBA design offers superior longitudinal phase space control without requiring additional elements, such as sextupole magnets, which could increase complexity and reduce system compactness. In addition, the modification of the angle of the center dipole magnet inside the TBA and quadrupole strength allows the flexible control of the R56 of the whole system. It could facilitate decompression or compression of the electron beam along with energy control, which will further enable the external seeded FEL schemes \cite{YANG2022167065,feng2022coherent, PhysRevLett.126.084801} and attosecond pulse generation \cite{Yan2024} in the CW-XFEL. Furthermore, the recently acheived superior RF performance of the 1.3 GHz superconducting cryomodule \cite{Chen2025} makes the multi beam-energy control scheme more feasible, paving the way for energy recovery linac operation in a CW-XFEL facility \cite{WANG2021165410}.

\vspace{10pt}
\noindent{\textbf{\ Acknowledgements}
\vspace{10pt}

This work was supported by the National Natural Science Foundation of China (12125508), the National Key Research and Development Program of China (2024YFA1612100), the CAS Project for Young Scientists in Basic Research (YSBR-042), and Shanghai Pilot Program for Basic Research – Chinese Academy of Sciences, Shanghai Branch (JCYJ-SHFY-2021-010).

\nocite{*}

\bibliography{reference}
        
\end{document}